\documentclass{article}
\usepackage[a4paper, margin=2cm]{geometry}
\usepackage{amsmath, graphicx}
\usepackage{xcolor}
\usepackage{transparent}
\usepackage{multirow}
\usepackage{tikz}
\usetikzlibrary{decorations.pathreplacing}
\usepackage{subcaption}
\usepackage{float}
\usepackage{hyperref}

\DeclareMathOperator{\SA}{SA}
\DeclareMathOperator{\CA}{CA}
\DeclareMathOperator{\MHA}{MHA}

\providecommand{\keywords}[1]{\textbf{\textit{Keywords ---}} #1}

\title{\textbf{Image Compression using only Attention based Neural Networks}}
\author{
\makeatletter
\parbox{0.25\linewidth}{\centering Natacha Luka\thanks{Thanks to \textit{Agence Innovation D\'{e}fense} (AID) and \'{E}cole des Ponts ParisTech for funding. This work was granted access to the HPC resources of IDRIS under the allocation 20XX-AD011011290 made by GENCI} \textsuperscript{1}\\
                \href{mailto:natacha.luka@enpc.fr}{natacha.luka@enpc.fr}}
    \and 
    \parbox{0.25\linewidth}{\centering Romain Negrel\textsuperscript{1}\\
    \href{mailto:romain.negrel@esiee.fr}{romain.negrel@esiee.fr}}
    \and 
    \parbox{0.25\linewidth}{\centering David Picard\textsuperscript{1}\\
    \href{mailto:david.picard@enpc.fr}{david.picard@enpc.fr}}
\and
\textsuperscript{1}LIGM, \'{E}cole des Ponts, Université Gustave Eiffel, ESIEE Paris,\and CNRS, Marne-La-Vallée, France
\makeatother
}
\date{}

\begin{document}

\maketitle
\begin{abstract}
In recent research, Learned Image Compression has gained prominence for its capacity to outperform traditional handcrafted pipelines, especially at low bit-rates. While existing methods incorporate convolutional priors with occasional attention blocks to address long-range dependencies, recent advances in computer vision advocate for a transformative shift towards fully transformer-based architectures grounded in the attention mechanism. This paper investigates the feasibility of image compression exclusively using attention layers within our novel model, \emph{QPressFormer}. We introduce the concept of learned \emph{image queries} to aggregate patch information via cross-attention, followed by quantization and coding techniques. Through extensive evaluations, our work demonstrates competitive performance achieved by convolution-free architectures across the popular Kodak, DIV2K, and CLIC datasets.
\end{abstract}
\keywords{Learned Image Compression, Vision Transformers, Visual Attention}
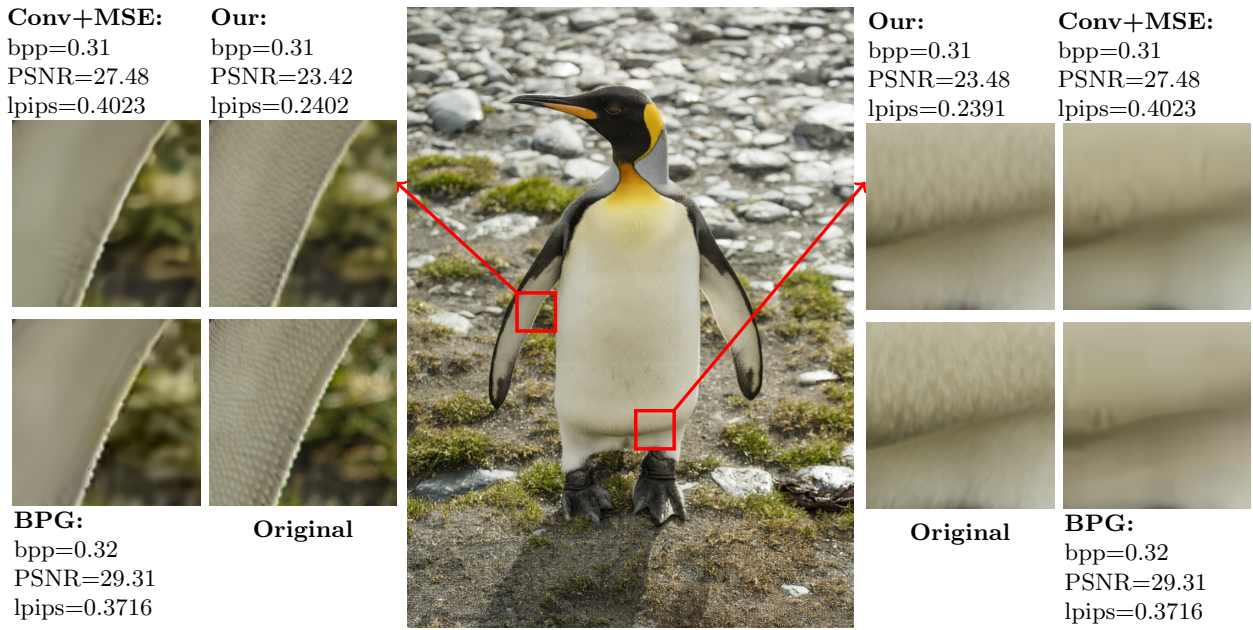
\begin{figure}
\small
\centering
    \begin{tikzpicture}[scale=0.925]
        \node[anchor=south west,inner sep=0] (image) at (0,0) {\includegraphics[scale=0.13, angle=-90]{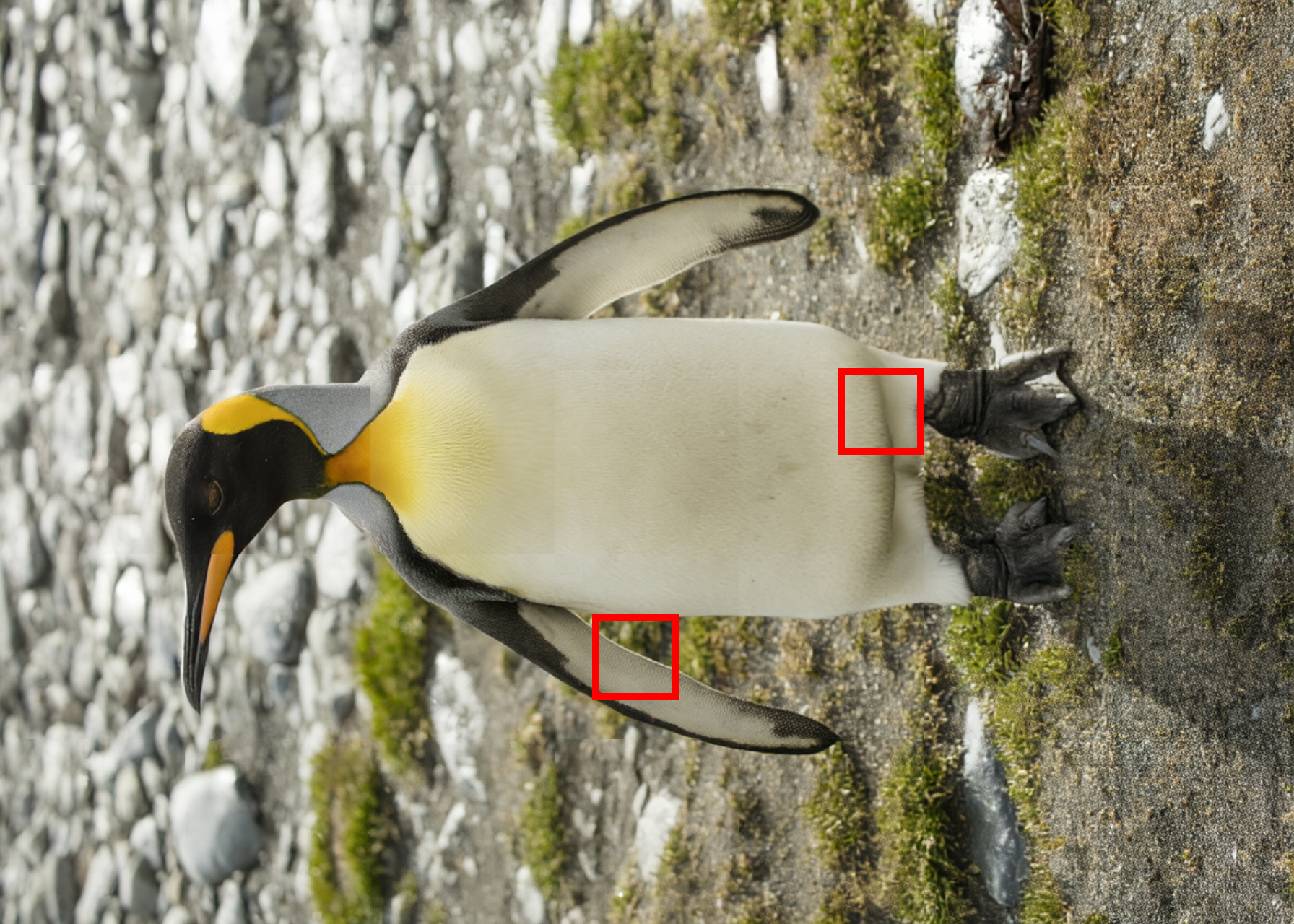}};
        \begin{scope}[x={(image.south east)},y={(image.north west)}]
            \draw[->=latex, very thick, color=red] (0.245,0.54) -- (-0.025,0.718) node[right] (patch1) {};
            \draw[->=latex, very thick, color=red] (0.597,0.35) -- (1.025,0.718) node[right] (patch2) {};
        \end{scope}
        
        \draw (patch1.east) ++(-1.6,-0.45) node (p1_our) {\includegraphics[scale=0.7, angle=-90]{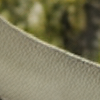}};
        \draw (p1_our) ++(-0.3,2.15) node (p1_our_txt) {
            \begin{tabular}{l}
                \textbf{Our:}\\
                bpp=0.31\\
                PSNR=23.42\\
                lpips=0.2402
            \end{tabular}};

        \draw (p1_our) ++(-2.8,0) node (p1_mse) {\includegraphics[scale=0.7, angle=-90]{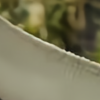}};
        \draw (p1_mse) ++(-0.3,2.15) node (p1_mse_txt) {
            \begin{tabular}{l}
                \textbf{Conv+MSE:}\\
                bpp=0.31\\
                PSNR=27.48\\
                lpips=0.4023
            \end{tabular}};
        
        \draw (patch1.east) ++(-1.6,-3.3) node (p1_or) {\includegraphics[scale=0.7, angle=-90]{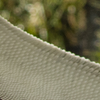}};
        \draw (p1_or) ++(-0,-1.7) node (p1_or_txt) {
            \textbf{Original}};

        \draw (p1_or) ++(-2.8,0) node (p1_bpg) {\includegraphics[scale=0.7, angle=-90]{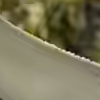}};
        \draw (p1_bpg) ++(-0.3,-2.2) node (p1_bpg_txt) {
            \begin{tabular}{l}
                \textbf{BPG:}\\
                bpp=0.32\\
                PSNR=29.31\\
                lpips=0.3716
            \end{tabular}};

        \draw (patch2.west) ++(1.35,-0.5) node (p2_our) {\includegraphics[scale=0.7, angle=-90]{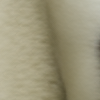}};
        \draw (p2_our) ++(-0.3,2.15) node (p2_our_txt) {
            \begin{tabular}{l}
                \textbf{Our:}\\
                bpp=0.31\\
                PSNR=23.48\\
                lpips=0.2391
            \end{tabular}};

        \draw (p2_our) ++(2.8,0) node (p2_mse) {\includegraphics[scale=0.7, angle=-90]{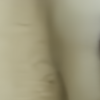}};
        \draw (p2_mse) ++(-0.3,2.15) node (p2_mse_txt) {
            \begin{tabular}{l}
                \textbf{Conv+MSE:}\\
                bpp=0.31\\
                PSNR=27.48\\
                lpips=0.4023
            \end{tabular}};

        \draw (patch2.west) ++(1.35,-3.35) node (p2_or) {\includegraphics[scale=0.7, angle=-90]{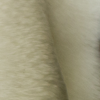}};
        \draw (p2_or) ++(0,-1.7) node (p2_or_txt) {
            \textbf{Original}};
                    
        \draw (p2_or) ++(2.8,0) node (p2_bpg) {\includegraphics[scale=0.7, angle=-90]{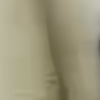}};
        \draw (p2_bpg) ++(-0.3,-2.2) node (p2_bpg_txt) {
            \begin{tabular}{l}
                \textbf{BPG:}\\
                bpp=0.32\\
                PSNR=29.31\\
                lpips=0.3716
            \end{tabular}};
    \end{tikzpicture}
    \caption{Image reconstruction from DIV2K. Zoom comparisons with original, convolutional equivalent model from \cite{balle2018variational_img_com_scale_hyperprior} and BPG are shown on the sides. Even at lower PSNR, our method significantly improves on the details of the penguin's fur, as captured by the lpips metrics. Examples on Kodak and CLIC in appendix \ref{apx:recons}.}
    \label{fig:img_div2k}
\end{figure}

\section{Introduction}
\label{sec:intro}

Image compression stands as a pivotal research domain, driven by the ubiquity of images in today's digital landscape. The persistent pursuit of higher image quality, while simultaneously optimizing bit size for storage and transmission, underscores its paramount importance. For many years, image compression codecs were meticulously handcrafted, yielding iconic solutions like the widely-used JPEG algorithm. In the post-JPEG era, a lineage of handcrafted codecs emerged, including JPEG2000 \cite{marcellin2002jpeg2000}, BPG format~\cite{bpg_format} (leveraging the HEVC video compression standard~\cite{sullivan2012overview_hevc_stdt}), and Google's WebP~\cite{webp_format}. Nevertheless, recent years have witnessed a transformative shift towards learned image compression, propelled by deep learning approaches. In the contemporary landscape, these learned codecs have not only achieved parity with their handcrafted counterparts but have also demonstrated the potential to surpass them.

In both hand-designed and learned image compression paradigms, common components are shared within the encoding and decoding processes. During encoding, an analysis transform maps the image from its pixel space representation to a more compression-friendly domain, followed by quantization, a pivotal step in lossy codecs. Subsequently, a lossless coding scheme translates the image data into a compact bitstream.
In the decoding phase, inverse operations unfold: the received bits are decoded, followed by synthesis through a suitable transform, ultimately restoring the image to its pixel space. A distinctive advantage inherent in learned compression methods arises from their holistic optimization approach. In this paradigm, all constituent blocks are collectively learned and end-to-end optimized. This optimization strategy diligently works to minimize distortion while concurrently minimizing bit-rate consumption.
Conversely, traditional hand-designed methods require the separate optimization of individual blocks, which may not inherently synergize when used together. Furthermore, hand-designed approaches involve human-centric decisions in selecting the transform, introducing inherent bias. In stark contrast, learned methods offer greater flexibility, as the network autonomously adapts both the analysis and synthesis transforms, determining the optimal compromise between rate and quality within specified constraints.

In learned image compression, numerous approaches have emerged for the analysis stage. Remarkably, transformer-based architectures, introduced by Vaswani et al. \cite{Vaswani_attention_is_all_you_need}, have remained relatively unexplored, despite their pervasive adoption in Computer Vision where they have demonstrated their supremacy over CNNs in many computer vision tasks \cite{liu2021swin_trans}. It is interesting to note that while attention or transformer blocks find occasional integration, they are almost invariably utilized \textit{in conjunction with} other usual convolution blocks, in both the analysis and synthesis processes \cite{li2022variable_rate_deep_img_comp_with_vision_trans, lu2022transformer_based_img_comp}. Furthermore, a critical prerequisite for effective image compression is the reduction of spatial redundancy, and the transformer's attention mechanism, which aggregates related elements within a sequence, appears especially well-suited for this purpose. With this context in mind, we propose in this paper a comprehensive investigation of the potential benefits of relying \textit{exclusively} on the transformer's attention mechanism for image compression, thereby eliminating convolutional priors from the equation.

The remaining of this paper is structured as follows: First, we present recent achievements in learned image compression, as well as basics on attention mechanisms. Then, we present and explain our transformer-based analysis/synthesis model based on our \emph{image queries} concept. These \emph{image queries} are learned prototypes which aggregate information from image step by step thanks to transformer cross-attention blocks. Finally, we show and analyze results of the proposed architecture on several datatsets to give some insights about the inner working of our proposed model \emph{QPressFormer}, before we conclude.

\section{Related Work}
\label{sec:related_work}

\subsection{Learned Image Compression}
\label{sec:related_image_comp}

In the realm of image compression, conventional lossless and lossy codecs, typified by ubiquitous standards like JPEG and its successors such as JPEG2000 \cite{marcellin2002jpeg2000}, WebP \cite{webp_format}, and BPG \cite{bpg_format} 
, have long prevailed as the industry norm. These well-established codecs continue to be indispensable in contemporary applications, exemplified by the ongoing development of the VVC codec \cite{gary2020vvc} for video, with adaptions tailored for image compression. Nonetheless, the recent years have witnessed a burgeoning interest in deep learning-driven approaches for both lossless and near-lossless image compression \cite{mentzer2019practical_full_res_learned_lossless_img_comp}, marking a noteworthy departure from conventional methodologies.

Todericci et al. introduced an innovative approach using recurrent neural networks (RNNs) \cite{toderici2016variale_rate_img_comp_rnn,toderici2017full_res_img_comp_with_RNN} for variable-rate image compression. Their method involves multiple passes of the original image through the network, generating an $m$-bit representation at each step to control the final bit-rate. While they explored various scenarios, they did not incorporate attention mechanisms.

Another strategy entails training multiple networks for distinct bit-rates. Ballé et al. introduced an innovative approach \cite{balle2017end_to_end_opt_img_comp, balle2018variational_img_com_scale_hyperprior} centered on learning the probability distribution of images within a latent space, with a concurrent focus on minimizing both the distortion and the entropy of this learned distribution. As per Shannon's theory, entropy serves as a lower bound for code length, rendering it synonymous with bit-rate minimization. The delicate equilibrium between these two objectives delineates the distortion/bit-rate trade-off. Their methodology incorporates convolutional transforms, paired with normalization functions \cite{balle2016density_modeling_img_using_gdn} for image compression.
To enhance the estimation of the latent probability distribution, Ballé and colleagues introduced side information through latent variable models, where the variables obtained from the analysis transform are treated as observed variables. A prior, referred to as the hyper-prior, is learned on these variables, effectively serving as a 'prior of the prior' to better account for spatial dependencies within the latent space. The ensuing hyper-latent variables are subsequently learned, contributing to the distribution in the hyper-latent space whose entropy is minimized to minimize bit-rate of this side-information. These hyper-variables allowing the variances inference of the original latent variables modeled as zero-mean Gaussian variables.

In recent works \cite{minnen2018joint_autoregressive_hierachical_priors_learned_img_comp}, significant advancements were made in optimizing the image compression pipeline, with two notable steps. Causal consideration of context began to be incorporated to enhance the inference of parameters within the distribution of subsequent latent variables. 
Building upon this foundation, attention mechanisms were integrated to encompass contextual information, wherein latent variables were modeled as Gaussian mixture models (GMMs), with parameter inferences facilitated by hyper latent variables \cite{Cheng2020learned_img_compression_gmm_attn}. Concurrently, several works embraced attention mechanisms to more effectively account for contextual nuances, particularly concerning bit allocation. These mechanisms dynamically allocate more bits to regions of higher attention, signifying their increased complexity and importance \cite{chen2021img_comp_nonlocal_attn,li2018learning_conv_net_for_content-weihted_img_comp}. Notably, while transformers have occasionally served as attention modules within these architectures \cite{lu2022transformer_based_img_comp,li2022variable_rate_deep_img_comp_with_vision_trans}
, they have been employed alongside convolutional components, distinguishing our approach.

Finally, generative models have also been investigated~\cite{Agustsson2019gan_for_extreme_learned_imf_comp, mentzer2021high-fidelity_generative_img_comp, agustsson2023multi-realism_img_com_with_conditional_gen}.
This results in better perception quality image especially in low bit-rate setup even if it is sometimes not reflected by distortion metrics like PSNR or MS-SSIM due to the Perception-Distortion trade-off~\cite{blau2018perception_distrosion_tradeoff}. Similarly to other approaches, even when attention is used it is in addition to convolutions.

\subsection{Transformers}
\label{sec:related_transformers}
Transformers, initially introduced by Vaswani et al. in the domain of Natural Language Processing (NLP) for translation tasks \cite{Vaswani_attention_is_all_you_need}, have since permeated diverse NLP applications, from BERT~\cite{devlin2019bert} to LlaMA~\cite{touvron2023llama} models. Within Computer Vision (CV), Dosovitskiy et al. inaugurated the integration of transformers \cite{dosovitskiy2020vit}, which are now prevalent in various vision tasks encompassing object detection \cite{carion2020end_to_end_detr}, image synthesis \cite{esser2021taming_transformers_highres_img_synth}, or cross-modal endeavors such as Text-Image Retrieval and Visual Question Answering \cite{chen2020uniter}. Notably, transformers have supplanted traditional CNNs \cite{liu2021swin_trans}.

The transformer architecture relies on an attention mechanism, facilitating interactions between two distinct sequences of tokens. Its fundamental objective is to iteratively aggregate tokens from a sequence denoted as \emph{values}, organized in a matrix $V$, into another sequence known as \emph{queries}, stored in a matrix $Q$. This aggregation is guided by the similarity between elements in $Q$ and $V$. To achieve this, the \emph{values} are subject to $h$ projections, distributed across different sub-units termed \emph{heads}, yielding $h$ sets of \emph{keys}, amalgamated within a matrix $K$. For each set of \emph{keys}, a similarity score $s_{i,j}$ is computed using the dot product between $q_i$ and $k_j$, where $k_j$ corresponds to specific features of $v_j$. These similarity scores are gathered within a matrix, normalized through the $\mathrm{softmax}$ function, and designated as the attention matrix $A$. Thus, at each iteration $n$, $h$ attention matrices are computed using the following equation:
\begin{align}
    A^n &= \sigma\left(Q^n\left(K^n\right)^T\right)\label{eq:attn_mat}
\end{align}
Once the attention matrices are obtained, they are employed to compute a weighted sum of the \emph{values}, subsequently added to the \emph{queries}. Consequently, the \emph{queries} accumulate the most akin \emph{values} corresponding to $h$ specific features, as determined through the projection of \emph{values} into $h$ sets of \emph{keys}. This process is concisely captured in the following equation, where $[\dots]_h$ denotes the aggregation of the $h$ sub-units:
\begin{align}
    Q^{n+1} &= Q^n + \left[A^nV^n\right]_h = Q^n + \left[\left(\sum_k a^n_{ik}v^n_{kj}\right)_{i,j}\right]_h\label{eq_attn_mha}
\end{align}
The equation \ref{eq_attn_mha} correspond to the main module of transformer called Multi-Head-Attention denoted $\MHA$ in the following.
In order to preserve the dimension $d$ of the whole transformer, all elements in the $h$ sub-units are projected in a space of dimension $d_h = \frac{d}{h}$, which implies $ d \equiv 0 \left(mod\ h\right)$.

There are 2 different $\MHA$ modules. The Self-Attention module, $\SA\left[Q\right]$,  correspond to the case where  each \emph{query} aggregates itself with its most similar \emph{queries} according to particular features. In this case \emph{queries} and \emph{values} are identical.
The other case correspond to the Cross-Attention , $\CA\left[Q,V\right]$. Here, the set of \emph{queries} and \emph{values} are different. The sequence of \emph{queries} aggregates information from another sequence. More details are available in~\cite{Vaswani_attention_is_all_you_need}.

\section{Method}

\begin{figure}
    \centering
    \def\svgwidth{0.52\columnwidth}
    \input{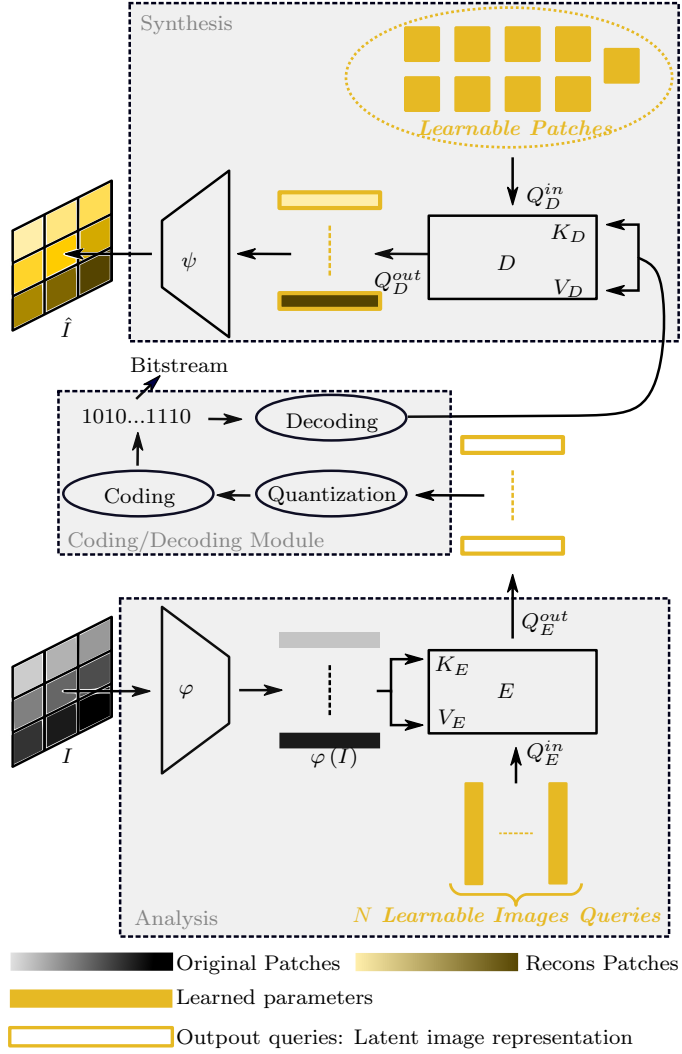}
    \caption{QPressFormer : Our Proposed Architecture.}
    \label{fig:arch}
\end{figure}

In this section, we introduce our convolution-free architecture, denoted QPressFormer, which exclusively relies on attention blocks. For the compression component, we adopt a factorized prior model similar to~\cite{balle2018variational_img_com_scale_hyperprior}. To ensure comparability, our aim is to develop a model with similar target bitrate, while replacing convolutional analysis and synthesis transforms with attention-based counterparts. The whole pipeline of our QPressFormer model is illustrated in Figure \ref{fig:arch}.

An image $I$ is first divided into patches, which are unrolled and augmented with positional encodings corresponding to their patch index using the transformation denoted as $\varphi$. The resulting set of vectors serves as the \emph{values}/\emph{keys} ($V_E$/$K_E$) for the encoder $E$, which employs attention blocks exclusively. Additionally, a set of $N$ learned vectors, referred to as 'Image queries' ($Q_E^{in}$), serves as \emph{queries}. The set $Q_E$ aggregates information from the image at each step $n$ through the attention process outlined in Section \ref{sec:related_transformers}. In-depth, the blocks of $E$ consist of transformer decoder blocks \cite{Vaswani_attention_is_all_you_need}, governed by the following equations for each layer $n$:
\begin{equation}
    \begin{aligned}
      Q^{n+1/2} &= \CA\left[\SA\left[Q^n\right] , \varphi(I)\right]\\
      Q^{n+1} &= Q^{n+1/2} + \mathrm{FFW}^n(Q^{n+1/2})
    \end{aligned}\label{eq:blk}
\end{equation}
At the final stage, 
the output \emph{queries} ($Q_E^{out}$) containing the compressed image. In this study, we work with 256x256 RGB images divided into 16x16 patches. This results in 256 patches, 
flattened as 768-dimensional vectors, which sets the embedded dimension of our architecture. We opt for $N=64$ \emph{Image Queries}. 
$Q_E^{out}$ corresponds to the latent variables so the probability distribution of $Q_E^{out}$ is estimated using a network and its entropy minimized to reduce the bit-rate, as discussed in Section \ref{sec:related_image_comp}. Importantly, our reconstruction perception metric is not the Mean-Square-Error (MSE) but lpips~\cite{Zhang_2018_CVPR_lpips}, a perceptual metric that aligns better with human perception.

%
The decoding process mirrors the encoding procedure. The vectors $\left\{Q_E^{out}\right\}$, containing the compressed image, serve as \emph{values} ($V_D$)/\emph{keys} ($K_D$) for the image decoder $D$. The \emph{queries} in $D$ are learned vectors corresponding to patch prototypes. They aggregate the compressed image progressively, and in the final stage, are reshaped and reassembled to yield the decompressed image $\hat{I}$.

The architecture is trained on the ImageNet using  Adam with a learning rate of 1e-4. Beyond 100,000 steps, optimizing lpips becomes more challenging. We observed that upscaling images to 512x512 before using lpips enhances results without requiring other changes. Both $E$ and $D$ have 12 heads and a depth of 12.

\section{Results}
    
\subsection{Quantitative Results}
In Table \ref{tab:res}, we present a comparative analysis between our model, BPG and equivalent convolutional factorized prior models from \cite{balle2018variational_img_com_scale_hyperprior} with pretrained weights obtained from CompressAI \cite{begaint2020compressai} on three datasets (Kodak~\cite{kodak_dset}, CLIC~\cite{CLIC2020} and DIV2K~\cite{div2k_dset}). For the MSE and MS-SSIM optimized models, we select weights corresponding to the closest bit per pixel (bpp) to our model. Our evaluation employs a range of metrics, including PSNR, MS-SSIM \cite{wang2003ms_ssim}, lpips \cite{Zhang_2018_CVPR_lpips}, and FID \cite{heusel2017fid}, which assesses the realism of generated images. 
Given that our architecture was trained on 256x256 images, we decompose all images into 256x256 segments and input each 256x256 part to the network before reassembling them. In cases where the image dimensions are not divisible by 256x256, we employ center cropping to obtain the largest segment divisible by 256x256. To ensure a fair comparison with convolutional counterparts or hand-designed BPG format, we apply the same image decomposition for reconstruction.

As shown in Table~\ref{tab:res}, the convolutional model optimized on MSE (respectively on MS-SSIM) performs better in terms of PSNR (respectively on MS-SSIM). In the same manner, our model performs significantly better on lpips. Moreover, our model performs better on FID even if it was not trained specifically for it. Besides, even if the PSNR and MS-SSIM of our model is lower than convolutionnal counterpart, it captures better perceptual details -- as expected with the lpips optimization -- as shown in figure \ref{fig:img_div2k}.
Convolution bias is thus not a prerequisite for Image compression and a transformer architecture seems at least equivalent or better -- in terms of perceptual metrics FID and lpips -- than convolutional networks and BPG.

\begin{table}
    \centering
        \begin{tabular}{l|c||c|c|c|c}
        Method  & bpp & PSNR$\uparrow$ & MS-SSIM$\uparrow$ & lpips$\downarrow$ & FID$\downarrow$\\
        \hline
        &&\multicolumn{4}{c}{\multirow{2}{*}{Kodak}}\\
        &&\multicolumn{4}{c}{}\\
        QpressFormer & 0.296 & 27.19 & 0.9203 & \textbf{0.2126} & \textbf{68.60} \\
        Conv+mse ~\cite{balle2018variational_img_com_scale_hyperprior} & 0.258 & 31.59 & 0.9608 & 0.3386 & 154.06 \\
        Conv+mssim ~\cite{balle2018variational_img_com_scale_hyperprior}& 0.320 & 30.66 & \textbf{0.9745} & 0.2913 & 117.94\\
        BPG~\cite{bpg_format} & 0.301 & \textbf{32.78} & 0.9626 & 0.2872 & 110.90 \\
        &&\multicolumn{4}{c}{\multirow{2}{*}{DIV2K (val-set)}}\\
        &&\multicolumn{4}{c}{}\\
        QpressFormer & 0.300 & 25.50 & 0.9159 & \textbf{0.2267} & \textbf{19.19} \\
        Conv+mse ~\cite{balle2018variational_img_com_scale_hyperprior}& 0.290 & 30.10 & 0.9642 & 0.3429 & 63.12 \\
        Conv+mssim ~\cite{balle2018variational_img_com_scale_hyperprior}& 0.322 & 28.94 & \textbf{0.9716} & 0.3237 & 49.78\\                BPG~\cite{bpg_format} & 0.307 & \textbf{30.84} & 0.9604 & 0.3155 & 54.73 \\
        &&\multicolumn{4}{c}{\multirow{2}{*}{CLIC (test-set)}}\\
        &&\multicolumn{4}{c}{}\\
        QpressFormer & 0.286 & 27.96 & 0.9339 & \textbf{0.2244} & \textbf{9.63} \\
        Conv+mse ~\cite{balle2018variational_img_com_scale_hyperprior}& 0.230 & 32.33 & 0.9657 & 0.3820 & 51.85 \\
        Conv+mssim ~\cite{balle2018variational_img_com_scale_hyperprior}& 0.277 & 31.32 & \textbf{0.9758} & 0.3536 & 36.83\\
        BPG~\cite{bpg_format} & 0.297 & \textbf{34.31} & 0.9712 & 0.3216 & 40.78
    \end{tabular}
    \caption{Quantitative results on Kodak, CLIC and DIV2K datasets.}
    \label{tab:res}
\end{table}

\subsection{Visual analysis}
    
\begin{figure}[h]
        \centering
        \begin{minipage}{0.6\linewidth}
        \includegraphics[width=\linewidth]{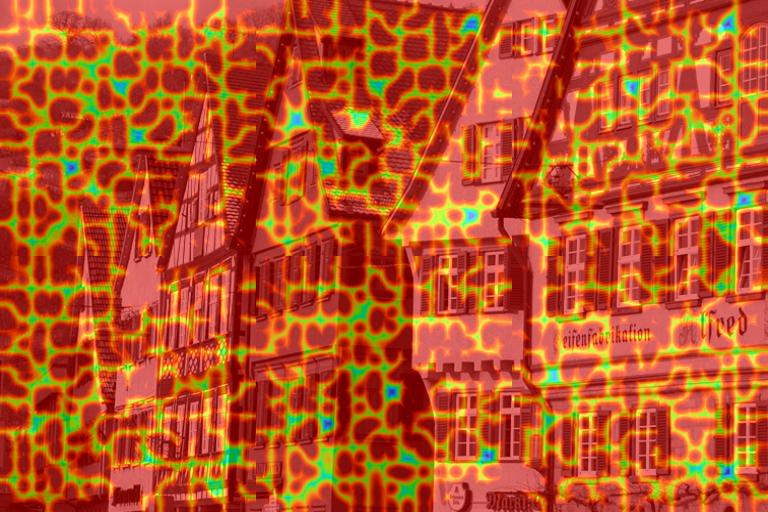}
        \includegraphics[width=\linewidth]{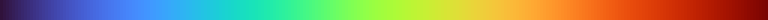}
        Low attn(0.015) \hfill High attn(1.000)
        \end{minipage}
        \caption{Attention map (max attn scores for all queries/layers/heads) for a Kodak image.}
        \label{fig:attn_colored_kodak}
    \end{figure}

To understand how the attention blocks process the image during encoding, we perform several visualizations. During the attention process, the attention weights are computed for a given pair of \emph{query} and \emph{key}. As each \emph{key} corresponds to a patch, we can see which part of the image one \emph{query} pays attention to by plotting its attention matrix. By combining the attention matrices of all queries, layers and heads (with \textbf{maximum}), we obtain a heatmap showing which parts of the image have been processed (Figure~\ref{fig:attn_colored_kodak}). As expected for good reconstruction, the whole image is attended, almost uniformly. In other words, at least one query pays attention only once in each image area.

\begin{figure}[h]
        \centering
        \begin{subcaptionblock}[t]{0.19\textwidth}
        \centering
        \includegraphics[width=\textwidth]{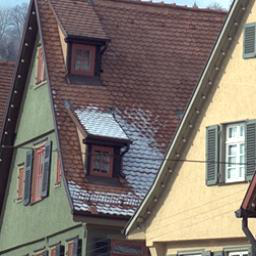}
        \caption{256$^2$ patch from Kodak}\label{fig:kodak_img_or}
        \end{subcaptionblock}
        \begin{subcaptionblock}[t]{0.19\textwidth}
        \centering
        \includegraphics[width=\textwidth]{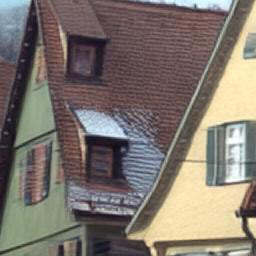}
        \caption{\centering Reconstruction}\label{fig:kodak_img_dec}
        \end{subcaptionblock}
        \begin{subcaptionblock}[t]{0.19\textwidth}
            \centering
            \begin{minipage}[b]{\textwidth}
                {\footnotesize Low(0.005) \hfill High(0.251)}
                \includegraphics[width=\textwidth]{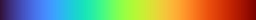}
                \includegraphics[width=\textwidth]{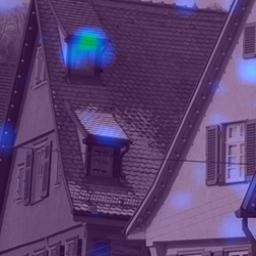}
            \end{minipage}
            \caption{Mean Attn of $\left\{Q_E\right\}_{11}$}\label{fig:kodak_attn}
        \end{subcaptionblock}
        \begin{subcaptionblock}[t]{0.19\textwidth}
            \includegraphics[width=\textwidth]{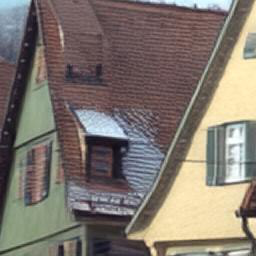}
            \caption{Reconstruction$^\dag$}\label{fig:kodak_no11}
        \end{subcaptionblock}
        \begin{subcaptionblock}[t]{0.19\textwidth}
        \centering
        \includegraphics[width=\textwidth]{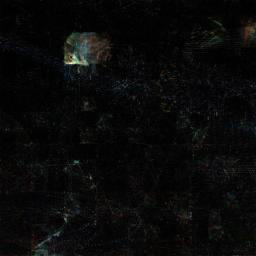}
        \caption{Mean recons. error$^\dag$}\label{fig:kodak_err_no11}
        \end{subcaptionblock}
    \caption{256x256 Kodak Image Crop Reconstruction with all $Q_E$ or with $\left\{Q_E\right\}_{11}$ removed (denoted $^\dag$).}
    \label{fig:Kodak_QE}
    \end{figure}

To better understand what information is captured by a \emph{query}, we visualize the layers and heads \textbf{mean} attention map for each \emph{query} individually. An example is given in Figure~\ref{fig:kodak_attn}. It is worth noting the \textbf{maximum} of attention maps and the \textbf{mean} of attention maps able us to make two distinct observations. The mean correspond to the area which is the most watched by the \textit{query} whereas the \textbf{maximum} allows us to ensure that all areas are watched by \textit{queries} at least one time. An original 256x256 crop from Kodak and the decoded patch are given in Figure~\ref{fig:kodak_img_or} and \ref{fig:kodak_img_dec}. The figure \ref{fig:kodak_attn} shows the \textbf{mean} attention map -- across heads and layers -- for \textit{query} 11. The attention is more focused on a part of the roof. To verify that this specific query encodes this particular region, we reconstruct the image without it. We can then see on the figure \ref{fig:kodak_no11} that the area where the removed \textit{query} paid attention to is not reconstructed properly. Interestingly, the area is inferred and in-painted with the roof texture, indicating that information from others patches able the encoder to infer information in that area.

In Figure~\ref{fig:kodak_err_no11}, the reconstruction error without this particular query and averaged over the Kodak dataset shows that the spatial information encode by the \emph{query} is the same for all images. In fact, after looking at the 64 averaged reconstruction error, nearly all \emph{queries} encode spatial information. It is worth noting that this spatial encoding property in the queries is an emergent behavior of the model, as it was not designed in the architecture or in the optimization.

However, spatial information is not the whole information encoded by \textit{queries} $Q_E^{out}$. First, a look at \textbf{maximum} attention maps of queries allow us to discover that the same \textit{query} pays attention to areas in different locations for different 256x256 crops of the same kodak image. Figure \ref{fig:appx_attn_map} in appendix \ref{appx:attn_map} gives more details. Second, if each \textit{query} was specialized spacially, in-painting wouldn't be possible.

\begin{figure}
        \centering
        \begin{subcaptionblock}[t]{0.49\textwidth}
        \centering
        \includegraphics[width=\textwidth]{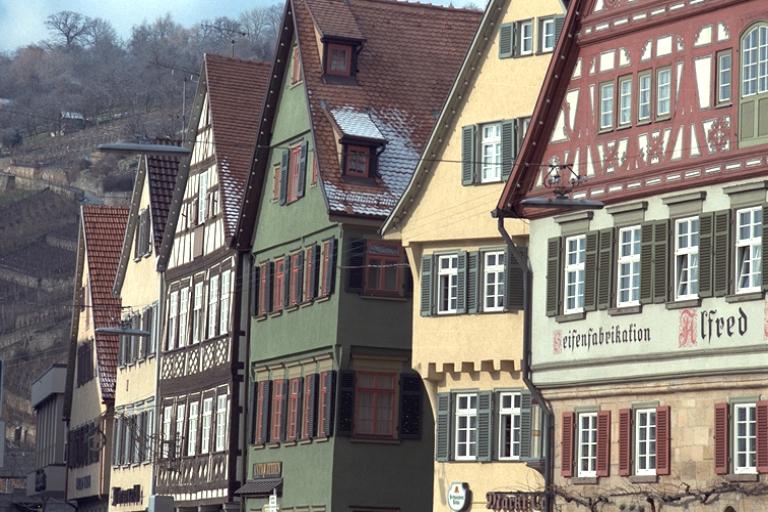}
        \caption{Kodak original Image}\label{fig:pca_or}
        \end{subcaptionblock}
        \begin{subcaptionblock}[t]{0.49\textwidth}
        \centering
        \includegraphics[width=\textwidth]{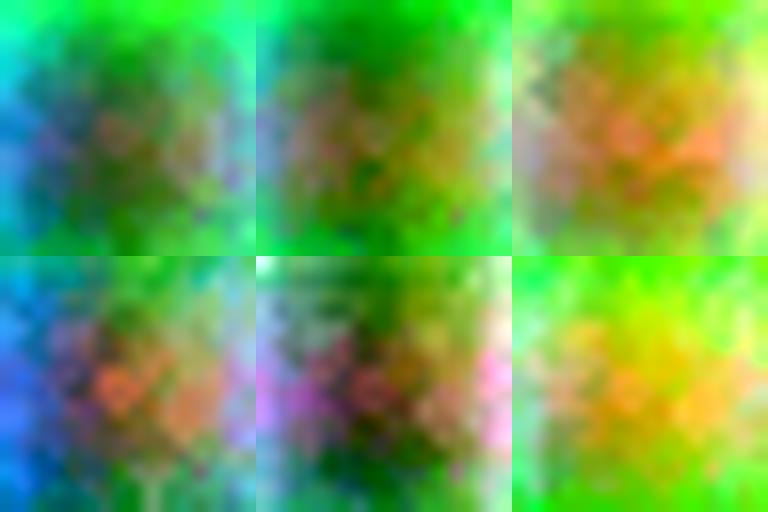}
        \caption{First layer of decoder cross-attention matrix projected on meta \textit{queries}}\label{fig:pca_spatial}
        \end{subcaptionblock}
        \begin{subcaptionblock}[t]{0.49\textwidth}
        \centering
        \includegraphics[width=\textwidth]{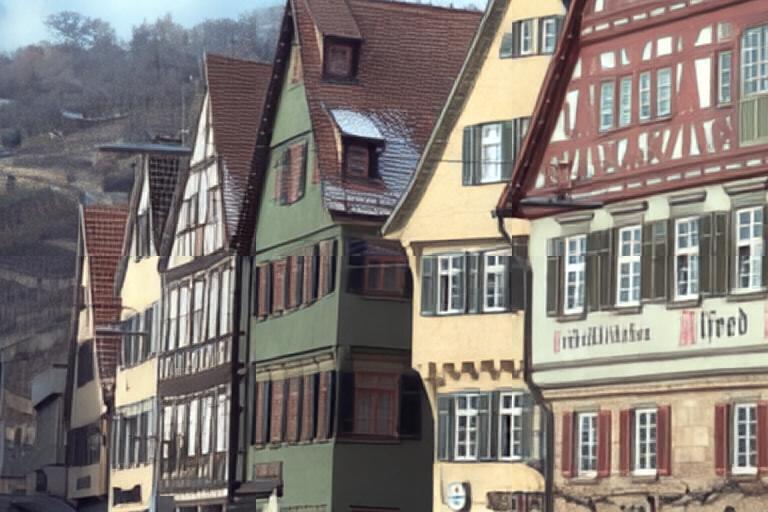}
        \caption{Decoded Kodak Image with \textit{QPressFormer}}\label{fig:pca_dec}
        \end{subcaptionblock}
        \begin{subcaptionblock}[t]{0.49\textwidth}
        \centering
        \includegraphics[width=\textwidth]{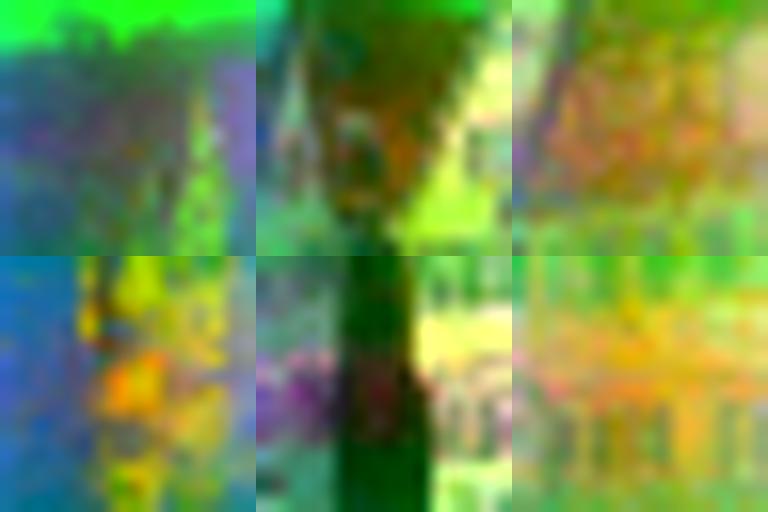}
        \caption{Tenth layer of decoder cross-attention matrix projected on meta \textit{queries}}\label{fig:pca_sem}
        \end{subcaptionblock}
\caption{Visualization of information encoded by output encoder \textit{queries} and extracted at layer 1 and 11 of the decoder. \ref{fig:pca_spatial} shows the layer 1 extract more spatial information from the \textit{queries} whereas \ref{fig:pca_sem} shows more semantics extraction with a nearly segmentation of image.}\label{fig:pca}
\end{figure}

To better visualize the elements encoded by \textit{queries}, we perform a PCA on the whole queries for a Kodak image to reduce the dimensions of \textit{queries} from 768 to 3. In the following we refers to these reduce dimensional \textit{queries} as \textit{meta-queries}. We can then visualize a RGB image corresponding to the main, or at least the most important, information decoded at each layer of the decoder.
Remember that the output \textit{queries} of the encoder $Q_E^{out}$ serve as \textit{keys/values} for the decoder wherin the \textit{queries} are learned patchs prototypes. After reducing the dimensionality of $Q_E^{out}$, we can project on it the mean across heads of cross-attention matrices (instead of the decoder \textit{values} obtained from the full encoder \textit{queries}) for a given layer of the decoder with the following equation:
\begin{align}
     I_{visu}^n &= \frac{1}{nb_{heads}}\sum_{h \in heads}\left(A_h^n\right)Q_{E;\left[:3\right]}^{out}\label{eq:pca}
\end{align}
In equation \ref{eq:pca} $Q_{E;\left[:3\right]}^{out}$ denotes the \textit{meta-queries} -- \textit{i.e.} output encoder \textit{queries} obtained after PCA for reducing dimension to 3. The figure \ref{fig:pca} shows clearly that \textit{queries} contain spatial information extracted at some specific layers: figure \ref{fig:pca_spatial} shows the same colors across different positions for different 256x256 blocks of an kodak image in addition to smooth transitioning between colors. In contrast figure \ref{fig:pca_sem} show a nearly segmentation of the original kodak image demonstrating some semantic information is extracted at others layers in the decoder, and so present in the queries. Thus, to perform a full reconstruction, \textit{queries} encodes at the same time semantic and spatial information. More visualizations are given in appendix \ref{apx:pca}

\section{Conclusion}
In this study, we explore learned image compression exclusively through attention blocks, eliminating convolutional biases. Our model, \emph{QPressFormer}, introduces "image queries" for cross-attention-based information aggregation. Our results show competitive results with convolutional counterparts with improved perceptual metrics. Future research should further investigate attention-based image compression. While we focused on analysis and synthesis components with a basic prior model, incorporating enhancements on prior model from convolutional models into attention-based architectures can assess transformer-based compression against state-of-the-art convolutional counterparts.

\clearpage
\bibliographystyle{IEEEbib}
\bibliography{refs}

\newpage
\begin{center} \Huge \textbf{Appendix}\end{center}
\normalsize
\appendix

\section{Reconstruction example on CLIC and Kodak}
\label{apx:recons}
\begin{figure}[H]
\small
\centering
    \begin{tikzpicture}[scale=0.925]
        \node[anchor=south west,inner sep=0] (image) at (0,0) {\includegraphics[scale=0.13]{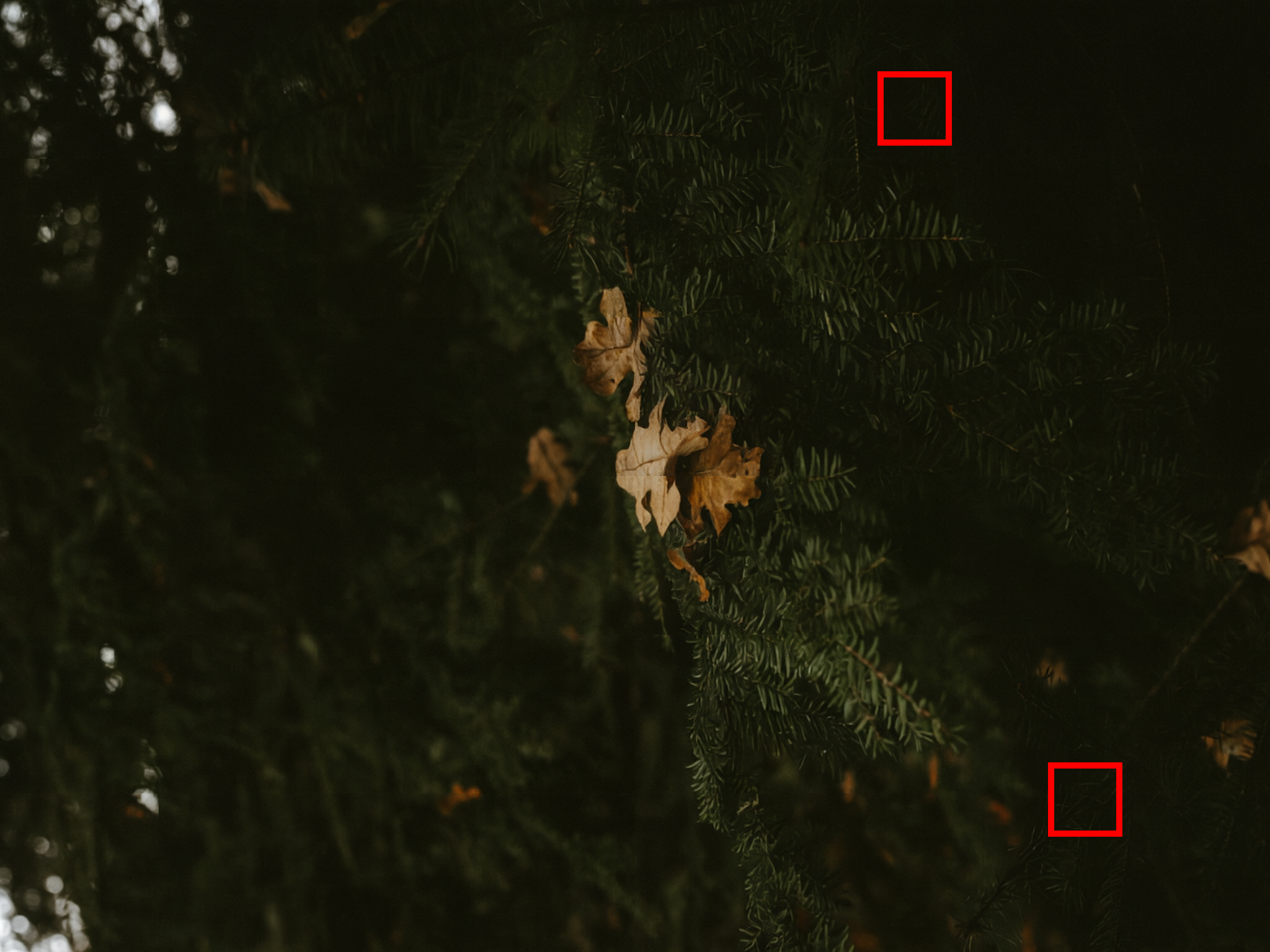}};
        \begin{scope}[x={(image.south east)},y={(image.north west)}]
            \draw[->=latex, very thick, color=red] (0.83,0.195) -- (-0.025,1) node[right] (patch1) {};
            \draw[->=latex, very thick, color=red] (0.745,0.92) -- (1.025,1) node[right] (patch2) {};
        \end{scope}
        
        \draw (patch1.east) ++(-1.6,-0.45) node (p1_our) {\includegraphics[scale=0.7]{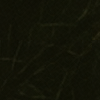}};
        \draw (p1_our) ++(-0.3,2.15) node (p1_our_txt) {
            \begin{tabular}{l}
                \textbf{Our:}\\
                bpp=0.26\\
                PSNR=33.07\\
                lpips=0.1230
            \end{tabular}};

        \draw (p1_our) ++(-0,-3.5) node (p1_or) {\includegraphics[scale=0.7]{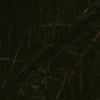}};
        \draw (p1_or) ++(-0,1.5) node (p1_or_txt) {
            \textbf{Original}};

        \draw (p1_or) ++(-0,-3.5) node (p1_mse) {\includegraphics[scale=0.7]{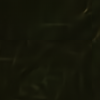}};
        \draw (p1_mse) ++(-0.2,-2.2) node (p1_mse_txt) {
            \begin{tabular}{l}
                \textbf{Conv+MSE:}\\
                bpp=0.29\\
                PSNR=38.46\\
                lpips=0.4682
            \end{tabular}};

        \draw (patch2.west) ++(1.35,-0.5) node (p2_our) {\includegraphics[scale=0.7]{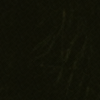}};
        \draw (p2_our) ++(-0.3,2.15) node (p2_our_txt) {
            \begin{tabular}{l}
                \textbf{Our:}\\
                bpp=0.26\\
                PSNR=33.07\\
                lpips=0.1230
            \end{tabular}};

        \draw (p2_our) ++(0,-3.5) node (p2_or) {\includegraphics[scale=0.7]{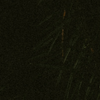}};
        \draw (p2_or) ++(0,1.5) node (p2_or_txt) {
            \textbf{Original}};

        \draw (p2_or) ++(0,-3.5) node (p2_mse) {\includegraphics[scale=0.7]{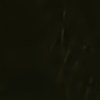}};
        \draw (p2_mse) ++(-0.2,-2.2) node (p2_mse_txt) {
            \begin{tabular}{l}
                \textbf{Conv+MSE:}\\
                bpp=0.29\\
                PSNR=38.46\\
                lpips=0.4682
            \end{tabular}};

    \end{tikzpicture}
    \caption{Image reconstruction from CLIC. Zoom comparisons with original and convolutional equivalent model from \cite{balle2018variational_img_com_scale_hyperprior} are shown on the sides. Even at lower PSNR, our method improves on the details of the leaves in dark areas.}
    \label{fig:img_clic}
\end{figure}

An example reconstruction on CLIC is given in figure \ref{fig:img_clic}.

\begin{figure}[H]
\small
\centering
    \begin{tikzpicture}[scale=0.925]
        \node[anchor=south west,inner sep=0] (image) at (0,0) {\includegraphics[scale=0.4]{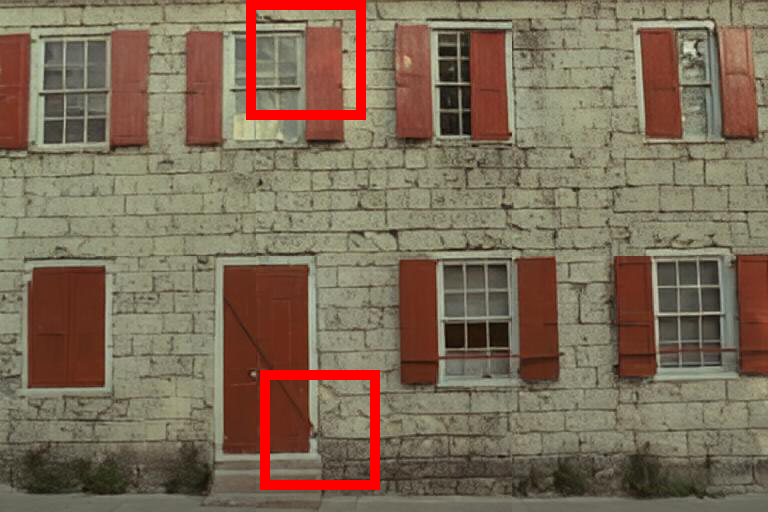}};
        \begin{scope}[x={(image.south east)},y={(image.north west)}]
            \draw[->=latex, very thick, color=red] (0.33,0.985) -- (-0.025,1.4) node[right] (patch1) {};
            \draw[->=latex, very thick, color=red] (0.345,0.276) -- (-.025,0.15) node[right] (patch2) {};
        \end{scope}
        
        \draw (patch1.east) ++(-1.6,-0.45) node (p1_our) {\includegraphics[scale=0.7]{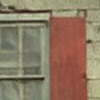}};
        \draw (p1_our) ++(-0.3,2.15) node (p1_our_txt) {
            \begin{tabular}{l}
                \textbf{Our:}\\
                bpp=0.30\\
                PSNR=25.16\\
                lpips=0.2217
            \end{tabular}};

        \draw (p1_our) ++(-2.8,0) node (p1_mse) {\includegraphics[scale=0.7]{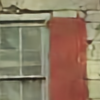}};
        \draw (p1_mse) ++(-0.4,2.15) node (p1_mse_txt) {
            \begin{tabular}{l}
                \textbf{Conv+MSE:}\\
                bpp=0.3461\\
                PSNR=28.37\\
                lpips=0.3161
            \end{tabular}};
        
        \draw (patch1.east) ++(-1.6,-3.3) node (p1_or) {\includegraphics[scale=0.7]{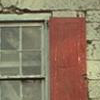}};
        \draw (p1_or) ++(-0,-1.7) node (p1_or_txt) {
            \textbf{Original}};

        \draw (p1_or) ++(-2.8,0) node (p1_ssim) {\includegraphics[scale=0.7]{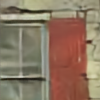}};
        \draw (p1_ssim) ++(-0,-2.2) node (p1_ssim_txt) {
            \begin{tabular}{l}
                \textbf{Conv+MS-SSIM:}\\
                bpp=0.28\\
                PSNR=25.92\\
                lpips=0.3473
            \end{tabular}};

        \draw (patch2.west) ++(-1.35,-0.5) node (p2_our) {\includegraphics[scale=0.7]{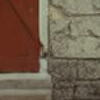}};
        \draw (p2_our) ++(-0.3,2.15) node (p2_our_txt) {
            \begin{tabular}{l}
                \textbf{Our:}\\
                bpp=0.30\\
                PSNR=25.16\\
                lpips=0.2217
            \end{tabular}};

        \draw (p2_our) ++(-2.8,0) node (p2_mse) {\includegraphics[scale=0.7]{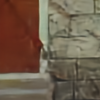}};
        \draw (p2_mse) ++(-0.3,2.15) node (p2_mse_txt) {
            \begin{tabular}{l}
                \textbf{Conv+MSE:}\\
                bpp=0.31\\
                PSNR=27.48\\
                lpips=0.4023
            \end{tabular}};

       \draw (patch2.west) ++(-1.35,-3.35) node (p2_or) {\includegraphics[scale=0.7]{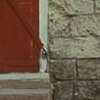}};
       \draw (p2_or) ++(0,-1.7) node (p2_or_txt) {
           \textbf{Original}};
                   
       \draw (p2_or) ++(-2.8,0) node (p2_msssim) {\includegraphics[scale=0.7]{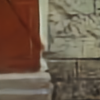}};
       \draw (p2_msssim) ++(-0.,-2.2) node (p2_msssim_txt) {
           \begin{tabular}{l}
               \textbf{Conv+MS-SSIM:}\\
                bpp=0.28\\
                PSNR=25.92\\
                lpips=0.3473
           \end{tabular}};
    \end{tikzpicture}
    \caption{Image reconstruction from Kodak. Zoom comparisons with original and convolutional equivalent models (optimised with MSE and MS-SSIM) from \cite{balle2018variational_img_com_scale_hyperprior} are shown on the sides.}
    \label{fig:img_kodak}
\end{figure}

\noindent An example reconstruction on Kodak is given in figure \ref{fig:img_kodak}.

\section{More PCA visualization}
\label{apx:pca}

\begin{figure}[H]
        \centering
        \begin{subcaptionblock}[t]{0.32\textwidth}
        \centering
        \includegraphics[width=\textwidth]{pca_dec/img_x_07.jpg}
        \end{subcaptionblock}
        \begin{subcaptionblock}[t]{0.32\textwidth}
        \centering
        \includegraphics[width=\textwidth]{pca_dec/attn_dec_img07_layer00_P1.jpg}
        \end{subcaptionblock}
        \begin{subcaptionblock}[t]{0.32\textwidth}
        \centering
        \includegraphics[width=\textwidth]{pca_dec/attn_dec_img07_layer10_P1.jpg}
        \end{subcaptionblock}
\begin{subcaptionblock}[t]{0.32\textwidth}
        \centering
        \includegraphics[width=\textwidth]{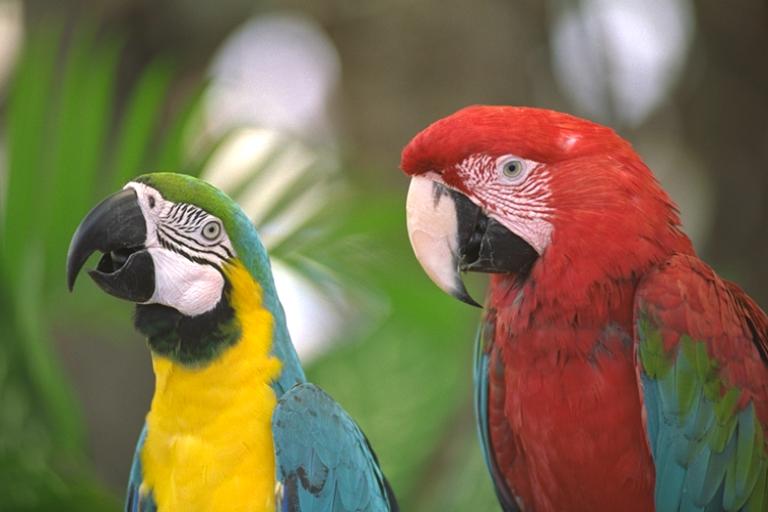}
        \end{subcaptionblock}
        \begin{subcaptionblock}[t]{0.32\textwidth}
        \centering
        \includegraphics[width=\textwidth]{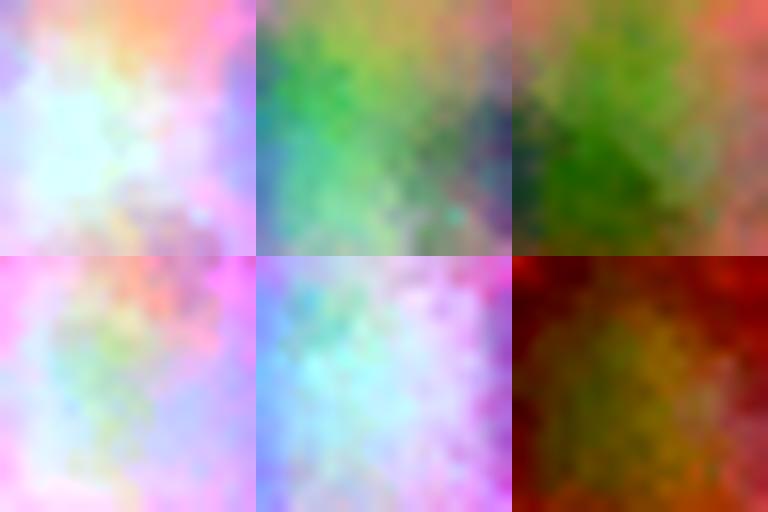}
        \end{subcaptionblock}
        \begin{subcaptionblock}[t]{0.32\textwidth}
        \centering
        \includegraphics[width=\textwidth]{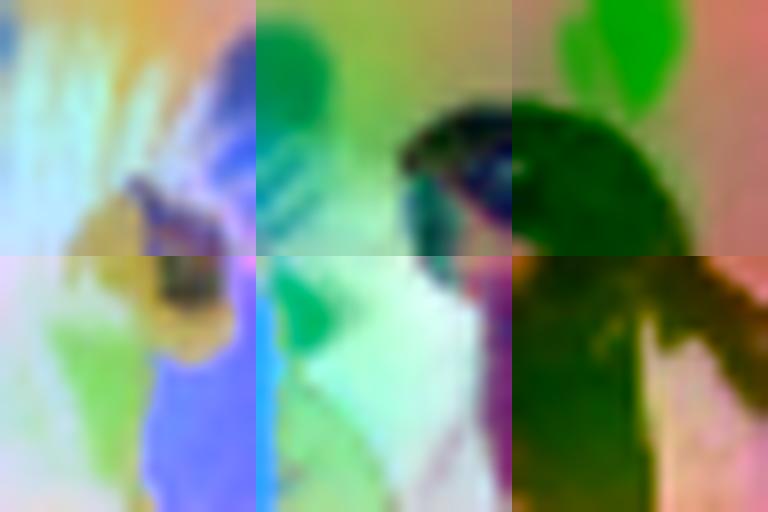}
        \end{subcaptionblock}
\begin{subcaptionblock}[t]{0.32\textwidth}
        \centering
        \includegraphics[width=\textwidth]{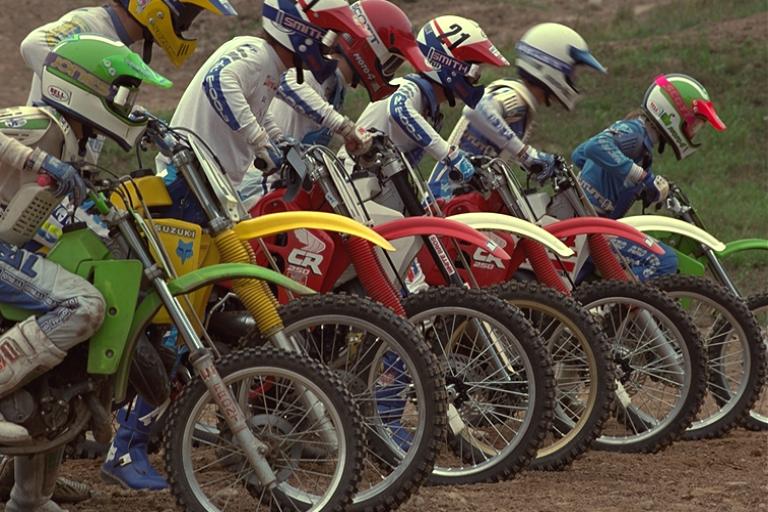}
        \end{subcaptionblock}
        \begin{subcaptionblock}[t]{0.32\textwidth}
        \centering
        \includegraphics[width=\textwidth]{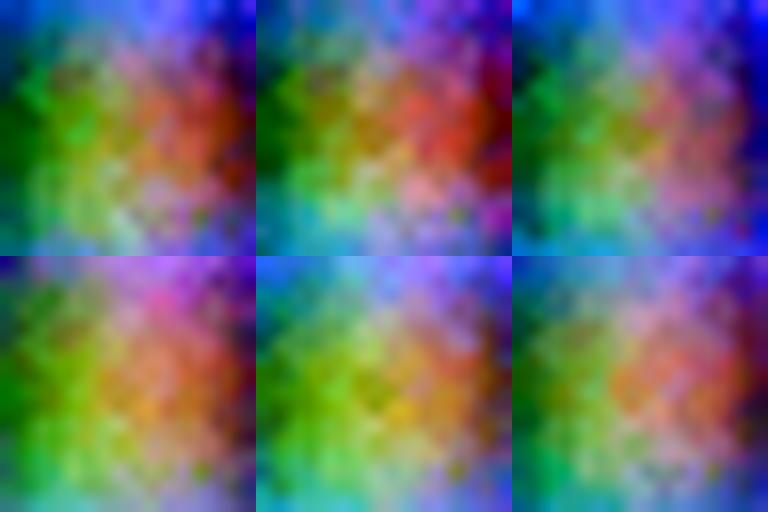}
        \end{subcaptionblock}
        \begin{subcaptionblock}[t]{0.32\textwidth}
        \centering
        \includegraphics[width=\textwidth]{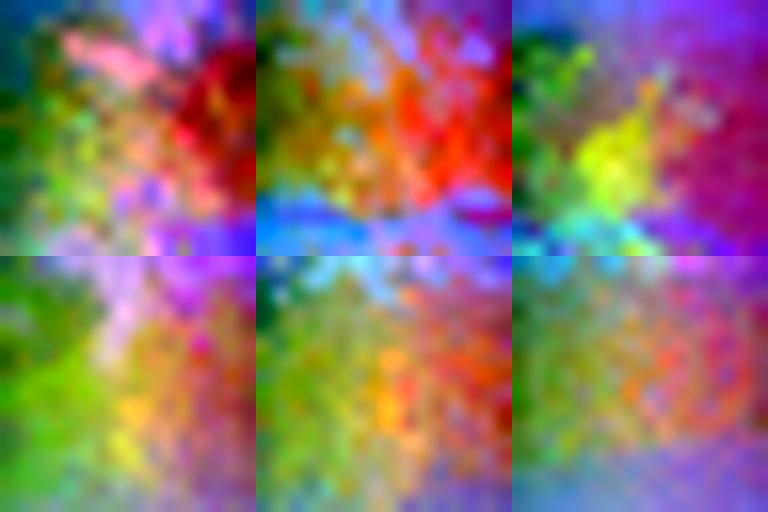}
        \end{subcaptionblock}
\begin{subcaptionblock}[t]{0.32\textwidth}
        \centering
        \includegraphics[width=\textwidth]{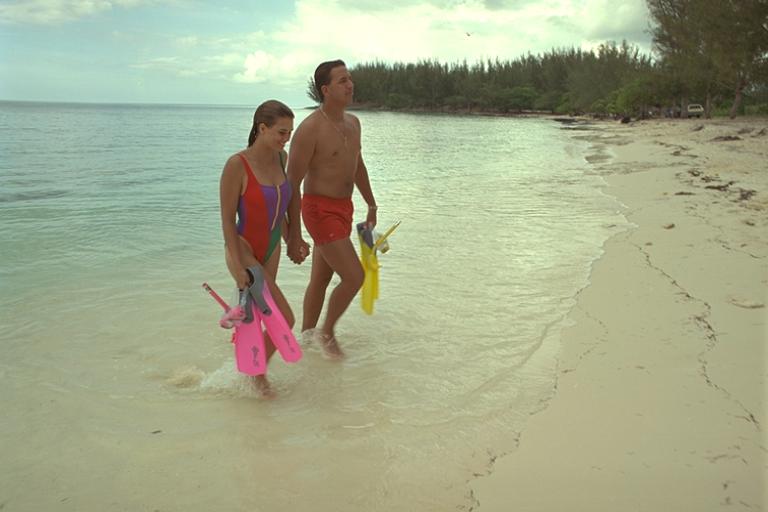}
        \end{subcaptionblock}
        \begin{subcaptionblock}[t]{0.32\textwidth}
        \centering
        \includegraphics[width=\textwidth]{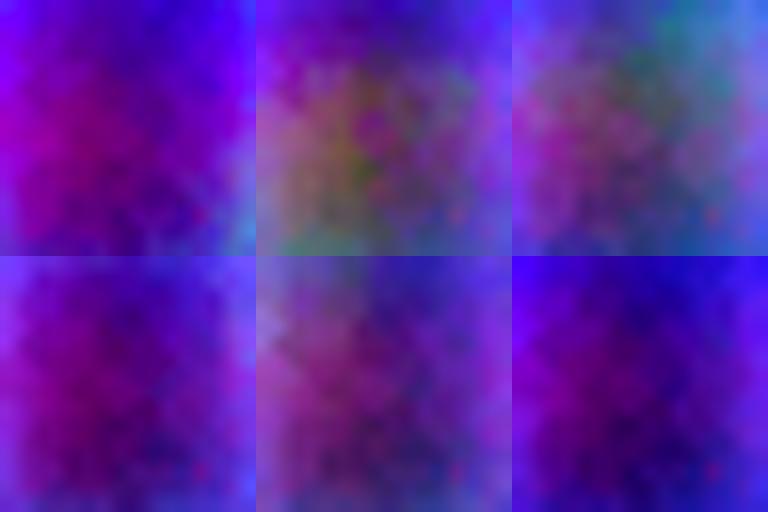}
        \end{subcaptionblock}
        \begin{subcaptionblock}[t]{0.32\textwidth}
        \centering
        \includegraphics[width=\textwidth]{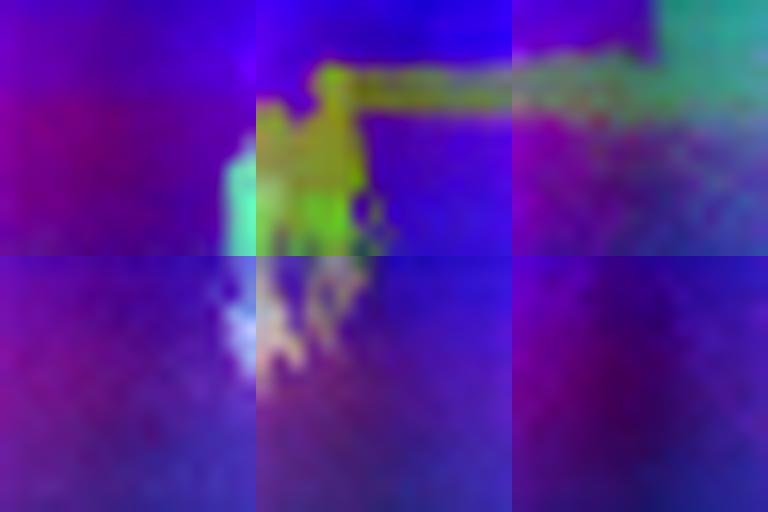}
        \end{subcaptionblock}
        \begin{subcaptionblock}[t]{0.32\textwidth}
        \centering
        \includegraphics[width=\textwidth]{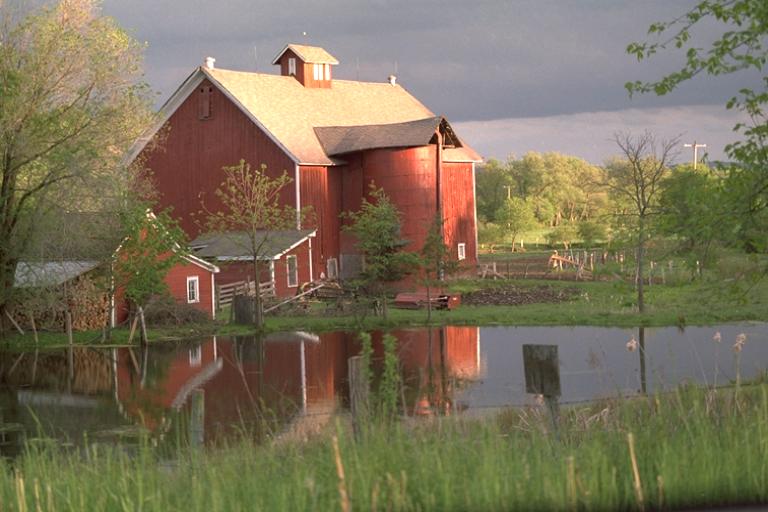}
        \caption{Column: Kodak original Image}\label{fig:appx_pca_or}
        \end{subcaptionblock}
        \begin{subcaptionblock}[t]{0.32\textwidth}
        \centering
        \includegraphics[width=\textwidth]{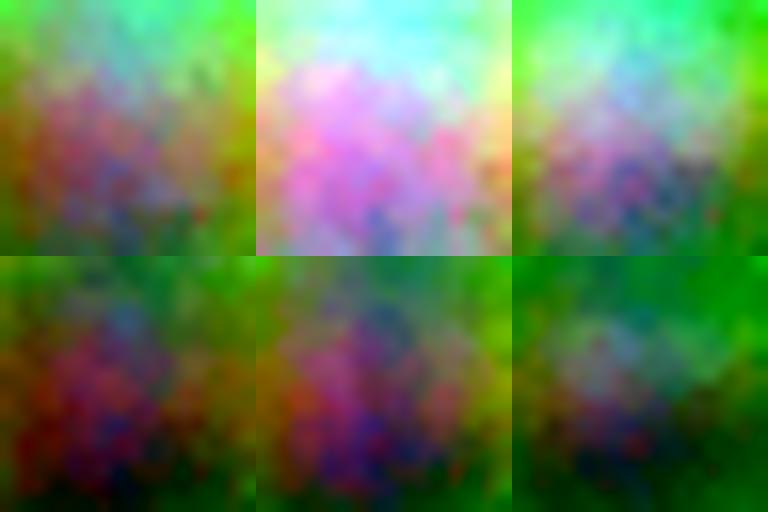}
        \caption{Column: $1^{st}$ layer of decoder cross-attention matrix projected on meta \textit{queries}}\label{fig:appx_pca_spatial}
        \end{subcaptionblock}
        \begin{subcaptionblock}[t]{0.32\textwidth}
        \centering
        \includegraphics[width=\textwidth]{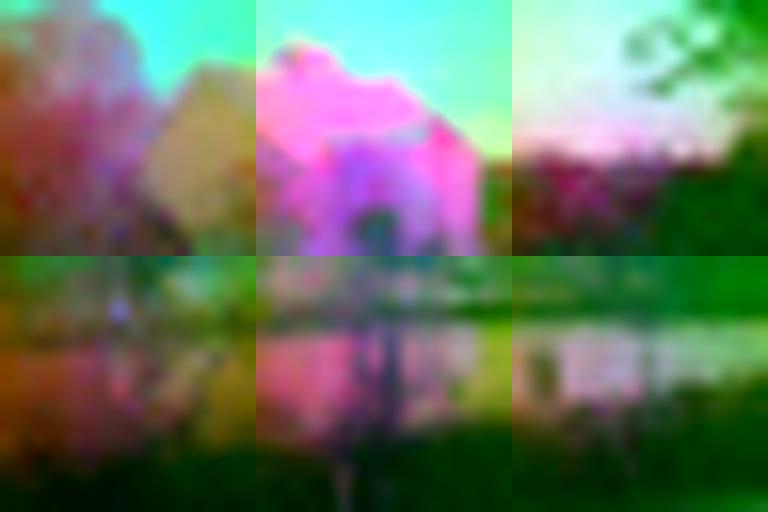}
        \caption{Column: $10^{th}$ layer of decoder cross-attention matrix projected on meta \textit{queries}}\label{fig:appx_pca_sem}
        \end{subcaptionblock}
\caption{Visualization of information encoded by output encoder \textit{queries} and extracted at layer 1 and 11 of the decoder. Column \subref{fig:appx_pca_spatial} shows the layer 1 extract more spatial information from the \textit{queries} whereas column \subref{fig:appx_pca_sem} shows more semantics extraction with a nearly segmentation of image.}\label{fig:appx_pca}
\end{figure}

Figure \ref{fig:appx_pca} shows that the first layer of the decoder extracts spatial information from the queries. Indeed by visualizing the projection of attention matrix on a YCbCr space\footnote{YCbCr follows by conversion in RGB gives better visualization colors} (then easily switched into RGB space) compute thanks to 3 first principal directions in the output encoder \textit{queries} space, we can see transitions between colors are smooth, with really similar or even same colors between all 256x256 crops of a same Kodak image (\textit{i.e.} with the same \textit{meta-queries} and so the same 3 directions used for constructing the RGB space). In an other layer of the decoder, semantics is extracted from \textit{queries} which able us to nearly segment the original image.   

\section{Maximum attention map vs mean attention map}
\label{appx:attn_map}

\begin{figure}[H]
        \centering
        \begin{subcaptionblock}[t]{0.49\textwidth}
        \centering
        \includegraphics[width=\textwidth]{pca_dec/img_x_07.jpg}
        \caption{Kodak original Image}
        \end{subcaptionblock}
        \begin{subcaptionblock}[t]{0.49\textwidth}
        \centering
        \includegraphics[width=\textwidth]{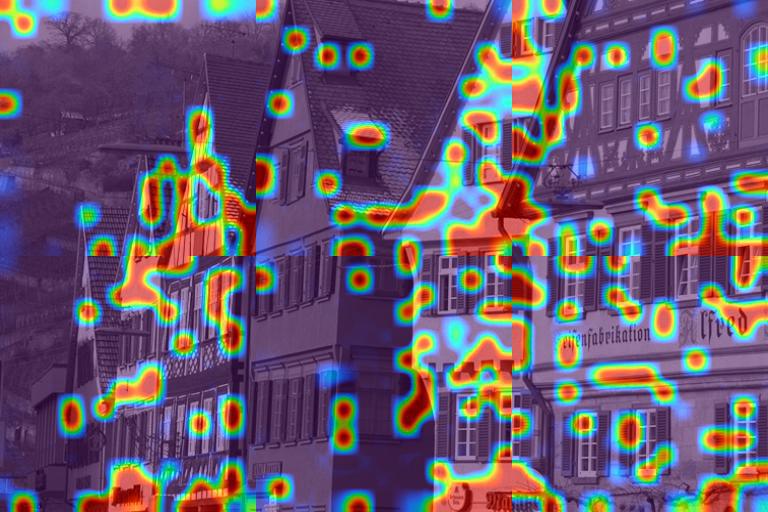}
        \caption{Maximum attention map of $\left\{Q_E\right\}_{36}$}\label{fig:appx_max_attn}
        \end{subcaptionblock}
        \begin{subcaptionblock}[t]{0.49\textwidth}
        \centering
        \includegraphics[width=\textwidth]{pca_dec/img_x_07_dec.jpg}
        \caption{Decoded Kodak Image with \textit{QPressFormer}}
        \end{subcaptionblock}
        \begin{subcaptionblock}[t]{0.49\textwidth}
        \centering
        \includegraphics[width=\textwidth]{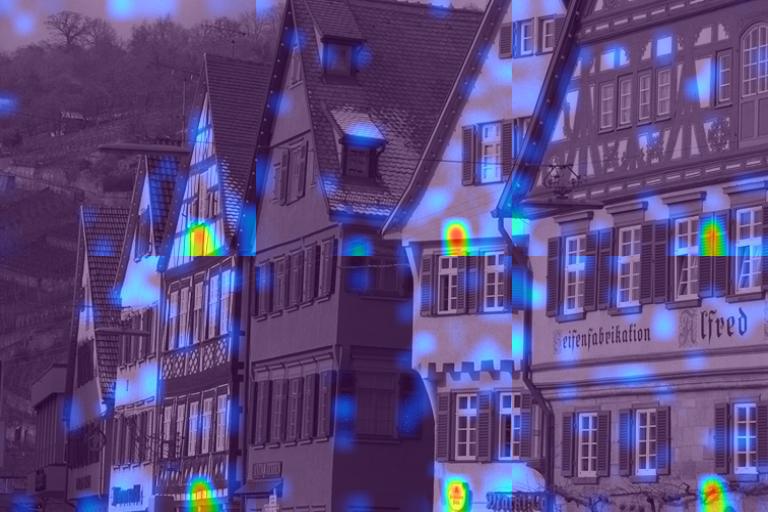}
        \caption{Mean attention map of $\left\{Q_E\right\}_{36}$}\label{fig:appx_mean_attn}
        \end{subcaptionblock}
        \begin{subcaptionblock}[t]{0.49\textwidth}
        \centering
        \includegraphics[width=\textwidth]{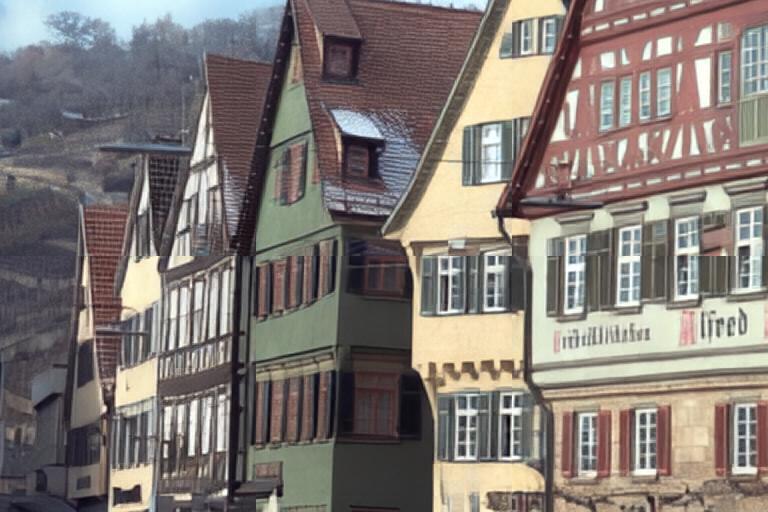}
        \caption{Image decoded without $\left\{Q_E\right\}_{36}$}\label{fig:appx_noq36}
        \end{subcaptionblock}
        \begin{subcaptionblock}[t]{0.49\textwidth}
        \centering
        \includegraphics[width=\textwidth]{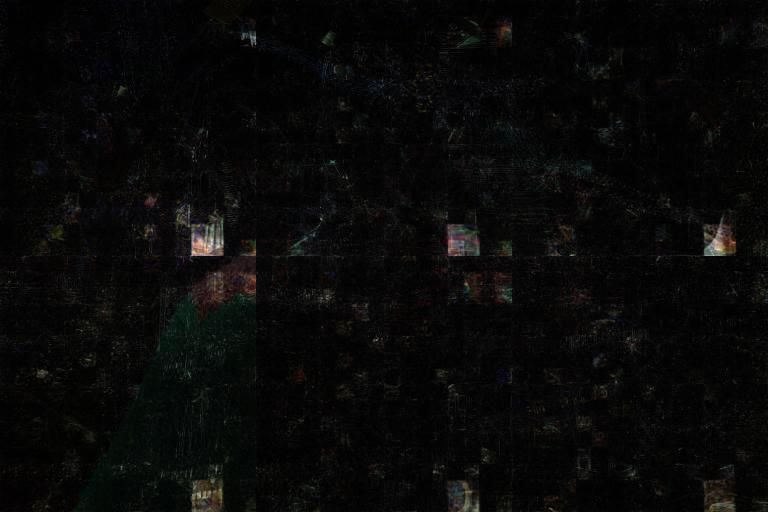}
        \caption{Mean Error of reconstruction without $\left\{Q_E\right\}_{36}$}\label{fig:appx_q36_err}
        \end{subcaptionblock}
    \caption{Difference between mean attention map and maximum attention map. Whereas mean attention map \ref{fig:appx_mean_attn} shows clearly the spatial encoding of the \textit{query} 36, the maximum attention map \ref{fig:appx_max_attn} is more difficult to interpret. But we can see that spatial information is not the only information encoded into \textit{queries} since different areas have high attention scores into different 265x256 blocks of the same image. \ref{fig:appx_noq36} and \ref{fig:appx_q36_err} corroborate the spatial encoding of \textit{queries}.}
    \label{fig:appx_attn_map}
\end{figure}

We show here the difference between mean attention maps -- \textit{i.e.} attention map obtained with the average of attention maps across heads and layers -- and the maximum attention maps -- \textit{i.e.} attention map obtained with the maximum of attention maps across heads and layers. In a case of mean attention maps, as many areas have nearly 0 attention somewhere in the encoding process, the mean below all attention values keeping only the ones which are high during the whole processing. We claim with this visualization we can see spatial encoding of \textit{queries}: attention weights of areas where a specific \textit{query} looks always during the encoding will not be reduced by the mean compared to other areas. In the opposite, the maximum attention maps allow us to see all areas where the \textit{query} looked at, but information encoded are more difficult to interpret, as shown on figure \ref{fig:appx_attn_map}.

\end{document}